\documentclass[aps,prb,reprint]{revtex4-1}
\usepackage{xcolor}
\usepackage[colorlinks=true,linktoc=page,citecolor=green]{hyperref}
\usepackage[all]{hypcap}
\usepackage{amsmath}
\usepackage{amssymb}
\usepackage{graphicx}
\usepackage{dcolumn}
\usepackage{bm}
\begin{document}

\title{Nonlocality in Homogeneous Superfluid Turbulence}
\author{O. M. Dix}
\author{R. J. Zieve}
\affiliation{Department of Physics, University of California Davis, 
Davis, California 95616, USA}
\begin{abstract}
Simulating superfluid turbulence using the localized induction approximation
allows neighboring parallel vortices to proliferate. In many circumstances a
turbulent tangle becomes unsustainable, degenerating into a series of parallel,
non-interacting vortex lines.  Calculating with the fully nonlocal Biot-Savart
law prevents this difficulty but also increases computation time. Here we use a
truncated Biot-Savart integral to investigate the effects of nonlocality on
homogeneous turbulence. We find that including the nonlocal interaction up to
roughly the spacing between nearest-neighbor vortex segments prevents the
parallel alignment from developing, yielding an accurate model of homogeneous
superfluid turbulence  with less computation time. 
\end{abstract}
\maketitle

Superfulid helium can be described by a two-fluid model as comprised of a normal
fluid part with velocity $v_n$ and a superfluid part with velocity $v_s$. At
small velocities, the superfluid exhibits remarkable properties such as the
ability to flow without dissipation. But above some critical velocity,
turbulence sets in and new interactions must be considered.  Schwarz
\cite{schwarz82} provided a major computational breakthrough in understanding
superfluid turbulence as a tangle of vortex filaments, an idea first suggested
by Feynman \cite{feynman55} and investigated  by Vinen \cite{vinen57c}. Each
vortex filament is an effectively one-dimensional curve around which superfluid
flows.  Since the superfluid is incompressible, $\nabla\cdot\vec{v}_s=0$, the
flow field due to these vortices is determined by the Biot-Savart law. 
Kelvin's theorem indicates that vortices move at approximately the local
superfluid velocity, meaning that the motion of each segment of the vortex is
determined by the positions of all vortices in the tangle.  An additional
interaction term between the superfluid vortex and the normal fluid gives us
the vortex equation of motion:
\begin{align}
	\dot{\vec{s}}(\xi,t) = &\vec{v}_s + \dot{\vec{s}}_{\mathrm{Biot}}+
	\alpha\hat{s}^{\,\prime}\times\left(\vec{v}_{ns}-
	\dot{\vec{s}}_{\mathrm{Biot}}\right) \notag \\
	&-\alpha'\hat{s}^{\,\prime}\times\left[\hat{s}^{\,\prime}\times
	\left(\vec{v}_{ns}-\dot{\vec{s}}_{\mathrm{Biot}}\right)\right],
	\label{eq:toruseqofmotion}\\
	\dot{\vec{s}}_{\mathrm{Biot}} = &\frac{\kappa}{4\pi}\int\frac{(\vec{s}_o-\vec{s})\times\mathrm{d}\vec{\xi}_o}{|\vec{s}_o-\vec{s}|^3}. \notag
\end{align}
Here, $\vec{s}(\xi,t)$ is the position along the vortex, parametrized by the
arclength, $\xi$. The circulation $\kappa$ is a fundamental constant for the
superfluid vortex and the integral runs over all vortices within the tangle.
The vector  $\hat{s}^{\,\prime}$ is the unit vector tangent to the vortex
filament, along which $\mathrm{d}\vec{\xi}_o$ also points. $\alpha$ and
$\alpha'$ are temperature-dependent parameters characterizing the interaction
between the normal fluid and the vortex core, and the quantity $\vec v_{ns}$ is
equal to $\vec v_n-\vec v_s$.

The Biot-Savart integral diverges as the source point $\vec{s}_o$ approaches
the field point $\vec{s}$, so the integral is split into a local part and
nonlocal part~\citep{arms65,schwarz85}.  The local part extends from the radius
of the filament core, $a_0$, to some arclength, $l_\pm$, away from the point of
interest. The remainder of the vortex system is included in the nonlocal
part of the integral,
\begin{align*}
	\dot{\vec{s}}_{\mathrm{Biot}}(\xi,t) =& 
	\frac{\kappa}{4\pi}\hat{s}^{\,\prime}
	\times\vec{s}^{\,\prime\prime}\ln\left(
	\frac{2(l_+l_-)^{1/2}}{e^{1/4}a_0}\right), \\
	&+\frac{\kappa}{4\pi}\int^\prime
	\frac{(\vec{s}_o-\vec{s})\times\mathrm{d}\vec{\xi}_o}
	{|\vec{s}_o-\vec{s}|^3}.
\end{align*}
Since the local term dominates, the nonlocal part has often been
ignored. This is called the localized induction approximation (LIA) or
the local approximation. When this is done, there is no objective best
cutoff for $l_\pm$ in the local term so the average radius of curvature,
$\bar{R}$, is used and the coefficient in front of the local term is
referred to as $\beta$, sometimes with a constant $c$ of order unity
included:
\begin{equation*}
	\beta=\frac{\kappa}{4\pi}\ln\left(\frac{\bar{R}}{ca_0}\right).
\end{equation*}

Vortex behavior is strongly influenced by vortex reconnections:  vortices can
approach and touch at a point, exchange heads and tails, then withdraw. Schwarz
\cite{schwarz85} investigated these events, arguing that the process could be
modeled by an instantaneous swap once two vortex segments approach within some
cutoff distance. When the LIA is employed, vortex reconnections are the only
nonlocal interaction involved in the simulation.

\section{\label{sec:motivationtorus}Motivation: Open-Orbit Vortices}

\begin{figure}
	\centering
        \includegraphics{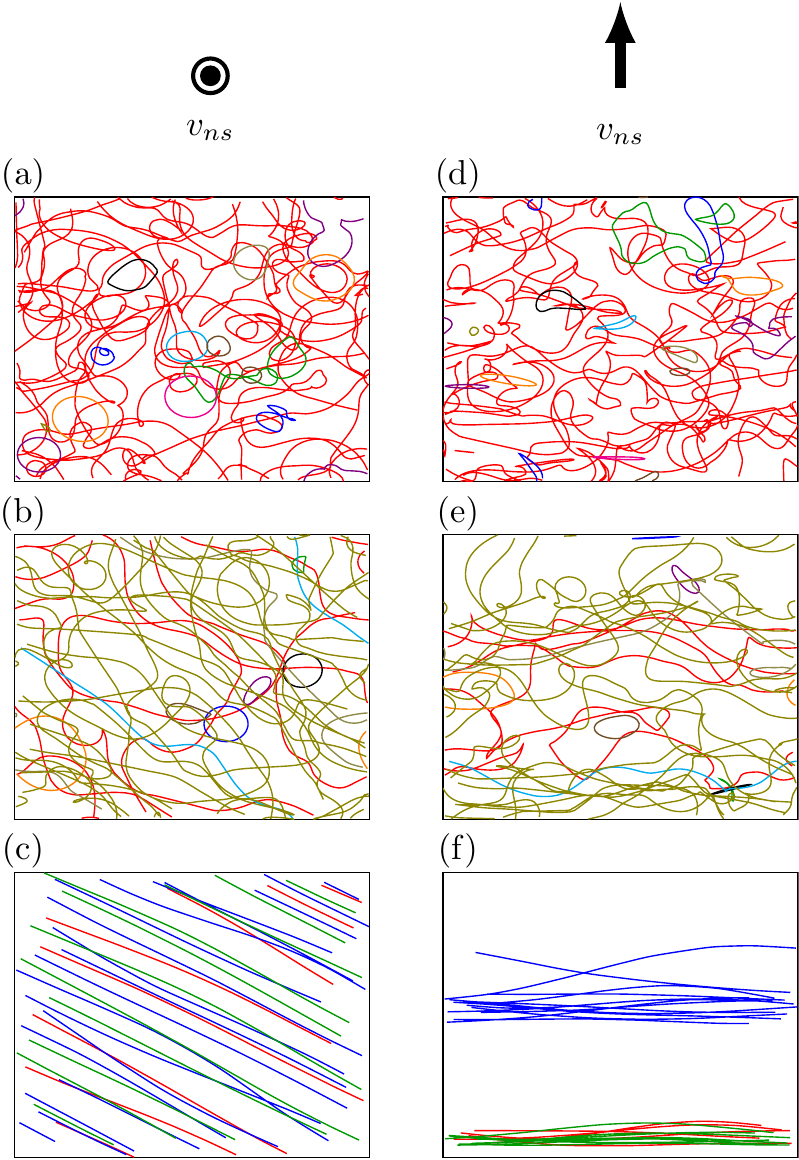}
	\caption{Projections of vortex tangles from the present work, with applied velocity perpendicular to (a-c) or within (d-f) the viewing plane. All computations use $v_{ns}=11$ cm/s, in a periodic cube with side length $50 \mu$m. The first row shows full Biot-Savart law calculation at time 0.142 s. The second and third rows show the LIA calculation at times 0.9 s and 0.142 s, respectively. In frames (b) and (e) the tangle has begun to exhibit anisotropy; in frames (c) and (f) the system has reached an open-orbit state, with the vortices almost entirely straight and parallel. 
	}
	\label{fig:vortextanglefullbiotvslia}
\end{figure}

Schwarz's early simulations \cite{schwarz85} apparently produced sustained
homogeneous turbulence. However, Buttke \cite{buttkecomment87} could only
reproduce the results by using deliberately inadequate spatial resolution.
Buttke subsequently argued \cite{buttke88} that since LIA does not allow vortex
stretching, it cannot describe superfluid turbulence. However, Schwarz
\cite{schwarzreply87} attributed the difference in computational results to an
artifact of the simulation geometry. With periodic boundary conditions, a vortex
line can cross the entire volume and close on itself. Energy loss will
increase the radius of curvature of such an ``open-orbit" vortex until the
vortex is entirely straight and interacts with the applied velocity field only
through an overall translation. No further energy is transfered between the
applied field and the vortex. The LIA contribution to the velocity of a straight
open-orbit vortex also vanishes, so only a reconnection can disrupt its
stability. In extreme cases the entire system can degenerate to an open-orbit
state, where all the vortices align and straighten into a clearly
non-homogeneous state that can persist indefinitely.
Figures~\ref{fig:vortextanglefullbiotvslia}c and
\ref{fig:vortextanglefullbiotvslia}f show such an open-orbit state.  To prevent
the open-orbit state from developing, Schwarz \cite{schwarz88} inserted an
occasional mixing step, in which half the vortices, selected at random, were
rotated by 90$^\circ$ about the direction of the applied flow. Buttke's
simulations, lacking this artificial mixing procedure, gave quite different
results.

Various other methods of avoiding the open-orbit state have been used.
Schwarz attributed the state to the periodic boundary conditions since it did
not occur when he used the much more computationally expensive real-wall
boundaries; however, the resulting tangles are not statistically identical to
those generated with periodic boundary conditions and a mixing
step\citep{schwarz88}. Aarts\cite{aartsdiss} does not include mixing, but
instead focuses on the time domain after the vortex line length equilibrates
but before the system degenerates into the open-orbit state.  Adachi {\em et
al.} \cite{adachi10} find that retaining the nonlocal terms instead of using
the LIA eliminates the need for any special accommodations to fix or avoid the
open-orbit state, suggesting that the LIA rather than the boundary conditions
may be at the root of the problem. Yet, as
Nemirovskii\cite{nemirovskiireview} notes, it is unclear why the nonlocal
term should be so effective, since nothing in the Biot-Savart integral
prevents the types of reconnections that produce open-orbit vortices.

Kondaurova {\em et al.}\cite{kondaurova08} suggest the vortex
reconnection condition as another possible source of trouble. Instead
of basing reconnection solely on the locations of neighboring vortices,
they consider the velocities of nearby vortex segments and carry out a
reconnection only if the segments would cross through each other during
the time step.  In a study comparing several reconnection algorithms,
Baggaley\cite{baggaley12recon} finds that the reconnection details have
little statistical effect on the vortex tangle in a nonlocal (Biot-Savart)
calculation.  However, results from LIA calculations do depend heavily
on when and how reconnections are carried out. Once again this hints at
a sickness in the LIA.

Here we propose yet another scheme for calculating the vortex motion between
reconnections. We include only those portions of the Biot-Savart integral where
$|\vec{s}_o-\vec{s}|$ is smaller than a nonlocal interaction distance $d_{NL}$,
and we vary $d_{NL}$ from 0 to interactions that span the
computational volume. Our results confirm that the LIA does not suffice for
modeling superfluid turbulence, but we find that only a small region of nonlocal
interaction is needed to reach accurate results. Our calculations also shed
light on some of the behaviors observed in previous computational work.

\section{\label{sec:reproducetorus}Results from
the Full Biot-Savart Integral and the LIA}

We first present data using the full Biot-Savart integral and the LIA to show
that our code is in good agreement with previous simulations and experiment. We
run simulations at temperature $T\approx1.6$~K, which corresponds to
$\alpha=0.1$. We ignore $\alpha'$ since it is an order of magnitude smaller,
and we use $v_n$ as our driving velocity ($v_s=0$) throughout this work to
eliminate uniform vortex translation. Except as otherwise noted, we use a
periodic cube with side length $D=50 \mu$m. Our equation of motion is
integrated using a Runge-Kutta-Fehlberg method (RKF54) with an adaptive time
step. Our point spacing is also adaptive, depending on the local radius of
curvature within the range $R/12\leq l\leq R/5$. As the vortex grows and a 
particular point spacing exceeds the upper limit, we add a new point along a
circular arc determined by the points that neighbor this overlarge spacing, as
described by Schwarz\cite{schwarz85}. In practice the time step is usually
about $2\times 10^{-6}$ s, and the spacing between points is
about $8\times 10^{-5}$ cm. We carry out a reconnection when vortices approach
within $2R_{min}/\ln(R_{min}/a_0)$, where $R_{min}$ is the smaller radius of
curvature at the points in question. The separation must also be below an
absolute cutoff, generally 10 $\mu$m. For our homogeneous vortex tangles, the
absolute level typically exceeds the curvature-based cutoff by more than an
order of magnitude and plays little role in the dynamics. As we discuss
below, the absolute cutoff becomes relevant in the untangled open-orbit state
produced with the LIA. We perform the reconnections themselves so as
not to increase the total vortex line length, by adjusting the position
along the vortex of the points involved in the reconnection\cite{owenthesis}.

We shall compare the LIA and full Biot-Savart calculations in multiple 
ways, including through visual 
inspection (as in Figure~\ref{fig:vortextanglefullbiotvslia}), using the 
line length density given by
\begin{equation}
	L = \frac{1}{V}\int\mathrm{d}\xi, \label{eq:ldens}
\end{equation}
where $V$ is the system volume, and with measures of the anisotropy 
given by~\citep{schwarz88}:
\begin{align}
	&I_{\parallel}=\frac{1}{V L}\int\left[1-\left(\hat{s}^{\,\prime}\cdot
\hat{r}_{\parallel}\right)^2\right]d\xi, \label{eq:Ipardef}\\
	&I_{\perp}=\frac{1}{V L}\int\left[1-\left(\hat{s}^{\,\prime}\cdot
\hat{r}_{\perp}\right)^2\right]d\xi, \label{eq:Iperpdef}\\
	&I_{\ell}\hat{r}_{\parallel}=\frac{1}{V L^{3/2}}\int\hat{s}^{\,\prime}
\times\vec{s}^{\,\prime\prime}d\xi. \label{eq:Ielldef}
\end{align}
Here $\hat{r}_{\parallel}$ and $\hat{r}_{\perp}$ are unit vectors parallel and 
perpendicular to the $\vec{v}_{ns}$ direction. 
As long as both directions perpendicular to the flow 
velocity are equivalent, the following relation is true 
regardless of tangle geometry: $I_{\parallel}/2+I_{\perp}=1$.
Note also that 
if a vortex tangle is completely isotropic, $I_{\parallel}=I_{\perp}=2/3$ and 
$I_{\ell}=0$. If vortices lie entirely within planes normal to $\vec{v}_{ns}$, 
$I_{\parallel}=1$, $I_{\perp}=1/2$, and $I_{\ell}$ will depend on the 
structure of the vortices. 

\begin{figure}
	\centering
	\includegraphics{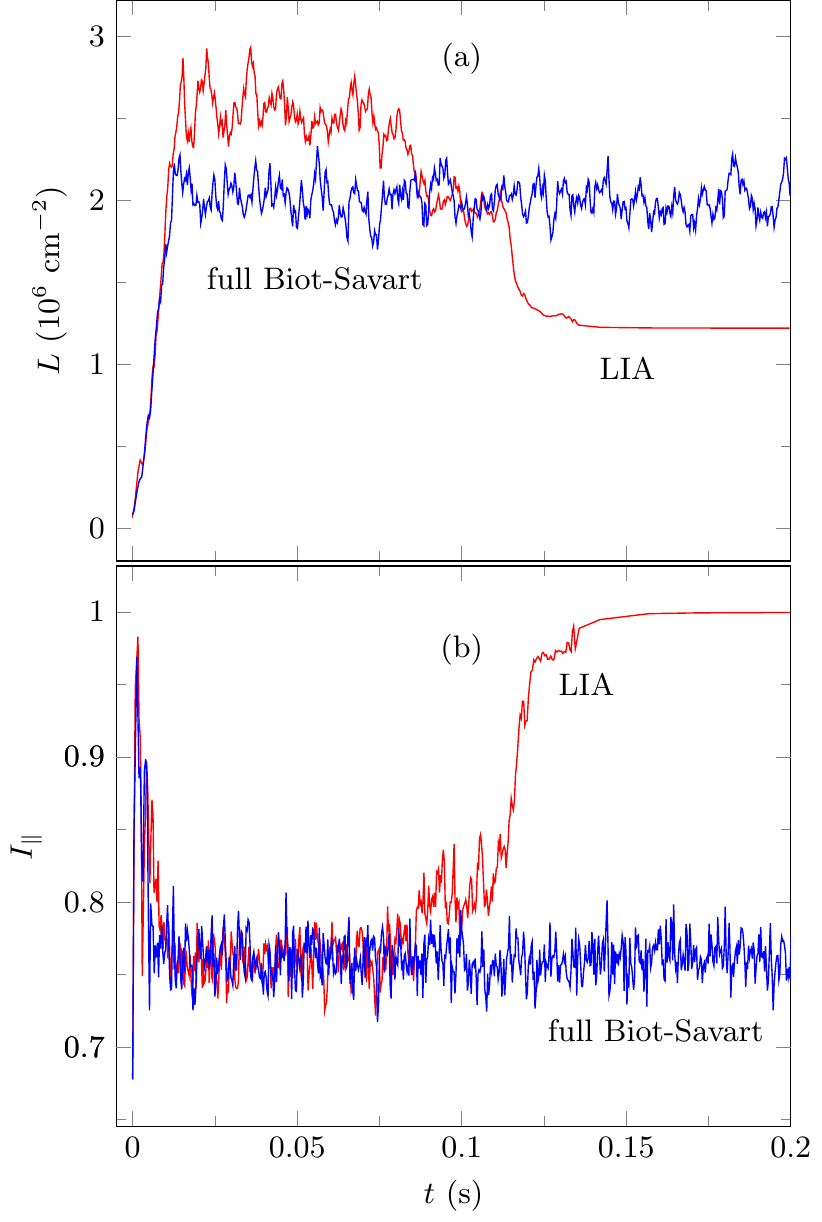}
	\caption{Results from LIA and full Biot-Savart 
	calculations for (a) $L(t)$ and (b) $I_\parallel(t)$, 
	with $v_{ns}=11$~cm/s. 
	}
	\label{fig:ldensfullbiotvsliatime}
\end{figure}

Figure~\ref{fig:ldensfullbiotvsliatime} compares 
$L(t)$ and $I_{\parallel}(t)$ for calculations with the 
LIA and full Biot-Savart law. The data agree well with those of
Adachi {\em et al.}\cite{adachi10}, with small differences arising from different system sizes and driving velocities.
Clearly the LIA results deviate significantly from those of the fully
nonlocal calculation. The presence of nonlocal interactions reduces
the vortex line length density.  As Adachi {\em et al.}\cite{adachi10}
explain, the nonlocal interaction is strongest immediately before and
after a reconnection, when vortices are closest.  The nonlocal term
tends to repel two parallel vortex segments and attract antiparallel
segments. Consequently, fewer parallel reconnections occur when the
nonlocal interaction is included. Furthermore, reconnections between
antiparallel segments produce highly curved regions that retreat away from
the reconnection site quickly. By contrast, reconnections of parallel
vortices result in more gently curved segments that do not retreat
quickly. The net result of these effects is that the nonlocal interaction
increases the average intervortex separation and correspondingly decreases
the line length density.

Figure \ref{fig:vortextanglefullbiotvslia} shows several snapshots
from these simulations. The first row is a homogeneous tangle from
the full Biot-Savart calculation. The LIA tangle appears similar for
$0.02<t<0.08$~s, a nearly steady-state regime. Aarts \cite{aartsdiss}
uses this regime to evaluate properties of the vortex
tangle. At later times, the system collapses into the open-orbit state,
with vortices aligning perpendicular to the driving velocity. The second
and third rows of Figure \ref{fig:vortextanglefullbiotvslia} illustrate
configurations during the collapse and after its completion.  How quickly
the open-orbit state forms depends strongly on the exact parameters used
for a simulation.  For example, it appears more quickly at high driving
velocities, where the growth of line length in the plane perpendicular to
the applied velocity helps to nucleate the open-orbit state.  In some
cases the intermediate, partially collapsed state can continue for
significant times, possibly indefinitely. The density of the open-orbit
state is an artificial value, determined by the absolute cutoff distance
used for reconnections. While this value plays almost no role for a highly
tangled state, it becomes the {\em de facto} reconnection distance as
vortices form open orbits and straighten. In the final state, neighboring
vortices are spaced far enough apart to avoid further reconnections.

\begin{figure}
	\centering
	\includegraphics{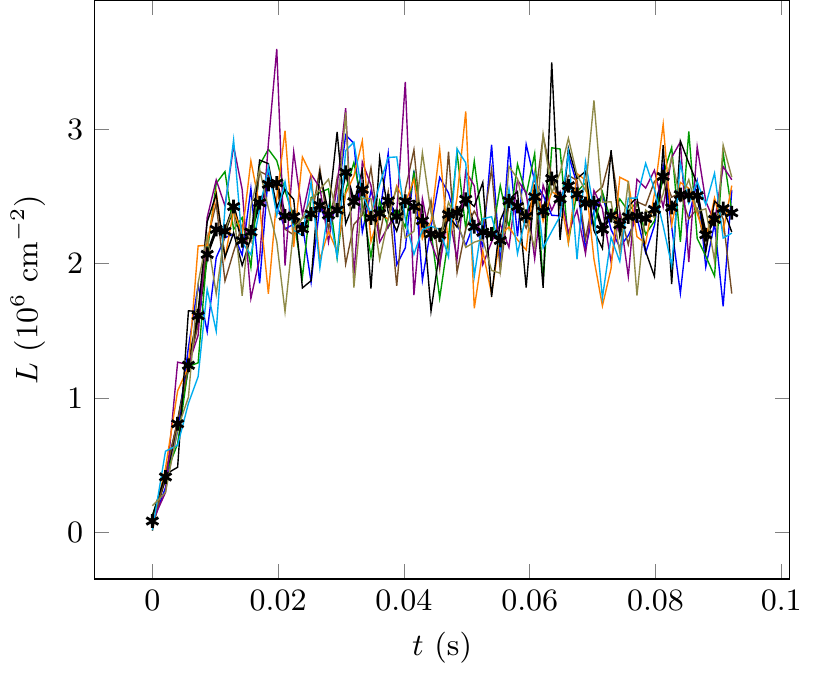}
	\caption{Evolution of the vortex line length density, $L(t)$, 
	within each octant of the periodic cube. The curve marked by black 		asterisks gives the line length per total system 
	volume, $D^3$. Each other curve gives $L(t)$ for 
	one of the eight $\mathbb{R}^3$ octants. The applied velocity is 
	$v_n=12$~cm/s.
	}
	\label{fig:homogldenstime}
\end{figure}

We demonstrate homogeneity of the full Biot-Savart calculation 
directly by finding the average line length 
density in different constituent volumes within the system 
(Equation~\ref{eq:ldens}). 
Figure~\ref{fig:homogldenstime} shows a sample evolution of $L(t)$. 
We use the eight $\mathbb{R}^3$ octants as the volumes to calculate each curve.

\begin{figure}
	\centering
	\includegraphics{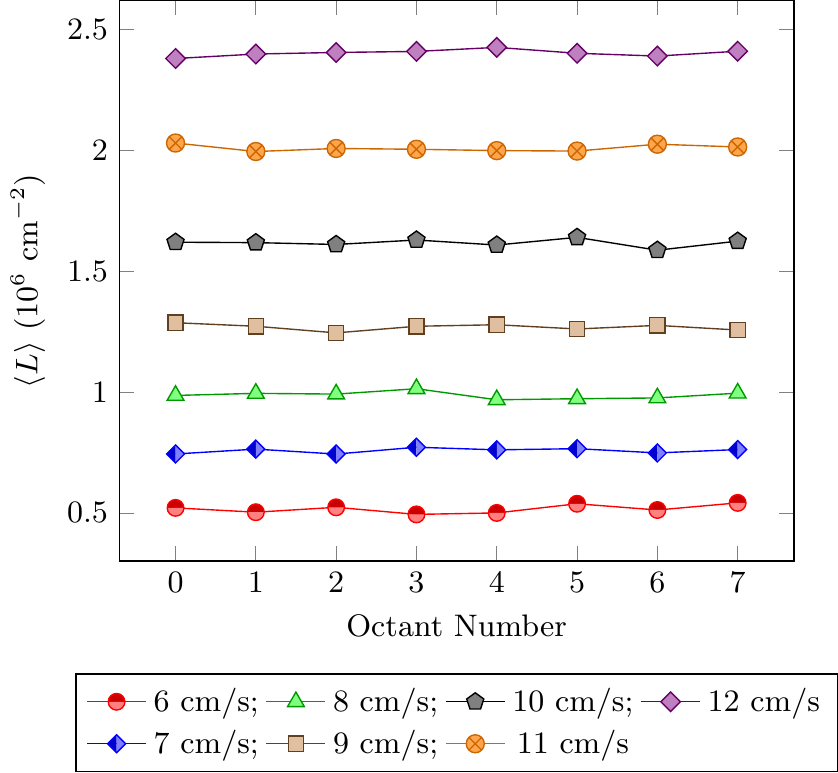}
	\caption{Average 
	equilibrated line length density, $\langle L\rangle$, 
	within each octant of the periodic cube, at multiple velocities. 
	}
	\label{fig:homogldens}
\end{figure}

Even after steady-state turbulence has set in, we expect some variation 
in $L(t)$ among the octants, but the average equilibrated values should 
be the same. Figure~\ref{fig:homogldens} shows these average 
equilibrated $L$ values for each octant, for a range of driving 
velocities. Time averages were done over the equilibrated time domain. 
We use the bracket notation $\langle X\rangle$ to denote the time average, 
as opposed to averages over the length of the vortex at a fixed time
for which we use the notation $\bar X$. From Figure~\ref{fig:homogldens}, it 
is clear that on this scale our system is homogeneous at every velocity. 

\begin{figure}
	\centering
	\includegraphics{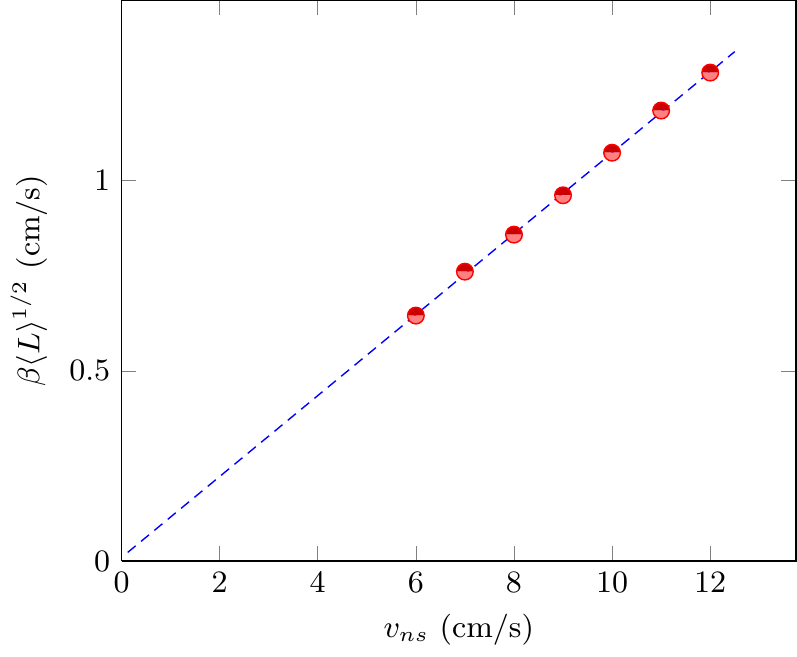}
	\caption{Average equilibrated line length density, $\langle L\rangle$, 
	for the total system volume, at multiple velocities. 
	The linear fit to our data yields a slope of $c_L=0.106$.
	}
	\label{fig:ldensbeta}
\end{figure}

Scaling arguments provide an additional test of homogeneous turbulence.
Simple dimensional analysis shows that the line length density
in homogeneous turbulence should depend on the driving velocity
as\citep{schwarz88}
\begin{equation}
	\langle L\rangle=c_L^2(v_{ns}/\beta)^2, \label{eq:ldensv}
\end{equation}
where $c_L$ is temperature-dependent. 
As noted previously, $\beta$ depends logarithmically on the average local 
radius of curvature, which decreases with increasing velocity. To 
keep $c_L$ independent of velocity, we do not combine $\beta$ with $c_L$.
As shown in Figure~\ref{fig:ldensbeta}, our simulations closely follow 
this scaling law. The best-fit value of $c_L$ is 0.106. 

\begin{figure}
	\centering
	\includegraphics{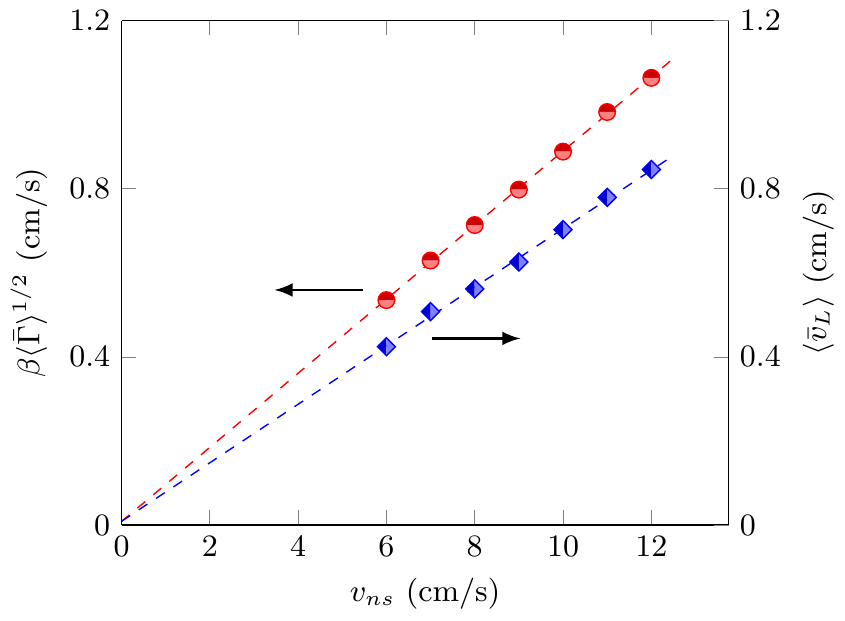}
	\caption{Left axis (red circles): 
	The quantity $\langle\bar{\Gamma}\rangle$ taken from the mutual 
	friction force density, $\bar{F}_{sn}$ along $\hat{v}_{ns}$. 
	The linear fit of this data yields a slope of $c_F=0.088$.
	Right axis (blue diamonds): The average vortex velocity, $\bar{v}_L$. 
	The linear fit of this data yields a slope of $c_v=0.070$.	
	}
	\label{fig:slopegammafric}
\end{figure}

Another scaling check of our simulations comes through the mutual friction 
force density due to interactions between the normal fluid and superfluid 
components. This force results in the interaction term of the vortex equation 
of motion ($\vec v_{fric}$), 
\begin{equation}
	\vec{F}_{sn}=-\frac{\rho_s\kappa}{V}\hat{s}^{\,\prime}\times\vec{v}_{fric},
		\label{eq:mutualfriction}
\end{equation}
where $V$ is the system volume. When averaged along the entire 
vortex, only $\bar{F}_{sn}$ parallel to the driving velocity is non-negligible. 
Schwarz \cite{schwarz88} also developed a scaling relation for the 
equilibrated force, $\langle\bar{F}_{sn}\rangle\propto v_{ns}^3/\beta^2$. 
We construct a new quantity:
\begin{equation}
	\bar{\Gamma} := \frac{\bar{F}_{sn}}{\rho_s\kappa\alpha v_{ns}},
	\label{eq:gammafric}
\end{equation}
which we can fit in the form 
$\beta\langle\bar{\Gamma}\rangle^{1/2}=c_F v_{ns}+y_0$, 
as shown in Figure~\ref{fig:slopegammafric}. The 
linear fit shows a slope of $c_F=0.088$ with no $y$-intercept, as 
theory predicts.

Yet another quantity used to describe homogeneous superfluid turbulence is 
the average vortex velocity relative to the applied superfluid velocity 
$\vec{v}_s$:
\begin{equation}
	\bar{v}_L=\left[\frac{1}{V L}\int \dot{s}\mathrm{d}\xi\right]-\vec{v}_s.
	\label{eq:avgvortexvelocity}
\end{equation}
As for $\bar{F}_{sn}$, only the component of $\bar v_L$ in the 
direction of the driving velocity is non-negligible. This time the quantity in 
question requires no scaling with $\beta$: $\bar{v}_L/v_{ns}=b(T)$, 
where $b(T)$ is a temperature-dependent parameter 
\citep{brewer82,schwarz88}. Our data for $\bar v_L$
are shown in Figure~\ref{fig:slopegammafric}.

The anisotropy parameters from Equations~\ref{eq:Ipardef} through
\ref{eq:Ielldef} are also consistent with previous simulations:
$\langle I_\parallel\rangle=0.761$, $\langle I_\perp\rangle=0.620$, 
$\langle I_\ell\rangle=0.646$.
These values show no significant trend over our range of $v_{ns}$. 
Schwarz \cite{schwarz88} points out that under the LIA,
the average mutual friction force density $\bar{F}_{sn}$ along
$\hat{v}_{ns}$ can be written as:
\begin{equation}
	\bar{F}_{sn}=\rho_s\kappa\alpha(c_L^2I_\parallel 
	- c_L^3I_\ell)v_{ns}^3/\beta^2. \label{eq:mutualfrictionanisotropy}
\end{equation}
As we have defined it, then: 
\begin{equation*}
	\bar{\Gamma}=(c_L^2I_\parallel - c_L^3I_\ell)
		\left(\frac{v_{ns}}{\beta}\right)^2.
\end{equation*}
We can see that the quantity $c_F$, obtained earlier 
from the mutual friction force density data, is equal to 
$(c_L^2 I_\parallel-c_L^3I_\ell)^{1/2}$.

Equation~\ref{eq:mutualfrictionanisotropy} is only approximate when nonlocal
contributions are included through the Biot-Savart law; hence we expect our
calculations to produce slightly different values from those of Schwarz
\cite{schwarz88} and Aarts \cite{aartsdiss}, who used the LIA. Additionally,
Schwarz \cite{schwarz88} states that, when neglecting $\alpha'$, the slope
parameter we call $c_v$ is equal to $c_L I_\ell$, again assuming the LIA. With
these relations, we compare characteristics of homogeneous turbulence in our
own simulations and in previous works, in Table~\ref{tab:homogturbquantities}.
We use $c_v$ or $c_L I_\ell$, depending on which can be derived from the data
shown in earlier works. Similarly, we use $c_F$ or $(c_L^2
I_\parallel-c_L^3 I_l)^{1/2}$. Schwarz \cite{schwarz88} does
show that his theoretical calculations of $I_\perp/I_\parallel$,
$(c_L^2I_\parallel-c_L^3I_\ell)^{1/2}$, and $c_LI_\ell$ match experiment.

\begin{table*}
	\centering
	\begin{tabular}{|c | c | c | c  c | c  c |}
	\hline
	Source (T=1.6~K) & $c_L$ & $I_\perp/I_\parallel$ & 
	$c_F$ & $(c_L^2I_\parallel - c_L^3I_\ell)^{1/2}$ & $c_v$ & $c_L I_\ell$\\
	\hline
	Present Work & 0.106 & 0.815 & 0.088 & 0.088 & 0.070 & 0.069 \\
	Aarts\cite{aartsdiss} & 0.11 & - & 0.095$^\dagger$ & - & 0.045$^\dagger$ 
		& - \\
	Schwarz\cite{schwarz88} & 0.137 & 0.775$^\ddagger$ & - & 0.116$^\dagger$ &	- &  0.063$^\dagger$ \\
	\hline
	\end{tabular}
	\caption{Comparison of 
	quantities characterizing homogeneous turbulence.
	$\dagger$: value calculated from other reported quantities; 
	$\ddagger$: value estimated from published figure.
	}
	\label{tab:homogturbquantities}
\end{table*}

Both Adachi {\em et al.}\cite{adachi10} and Kondaurova {\em et
al.}\cite{kondaurova08} report $\gamma=c_L/\langle\beta\rangle$
rather than $c_L$ itself.  Doing so requires one to insert an
ad hoc $y$-intercept to maintain a decent linear fit. When
we do this, we get a value of $\gamma=137.3$~s/cm$^2$,
where  Adachi {\em et al.}\cite{adachi10} report a value of
$\gamma=109.6$~s/cm$^2$. The experimental value is $\gamma=93$~s/cm$^2$
~\citep{childers76,brewer82}. We can extract the value for
$I_\perp/I_\parallel\approx0.8$ from Adachi {\em et al.}\cite{adachi10},
which also matches our quantity from Table~\ref{tab:homogturbquantities}
of $0.815$. Kondaurova {\em et al.}\cite{kondaurova08} find
$\gamma=280$~s/cm$^2$, which is extremely high. In more recent work by
the same lead author\cite{kondaurova14}, $\gamma$ ranges from 105.8 to
120.2 s/cm$^2$ at 1.6 K, depending on the reconnection criterion used. The
difference may be from the larger sample volume and much lower applied
fluid velocities used for the later calculations.

\begin{figure}
	\centering
	\includegraphics{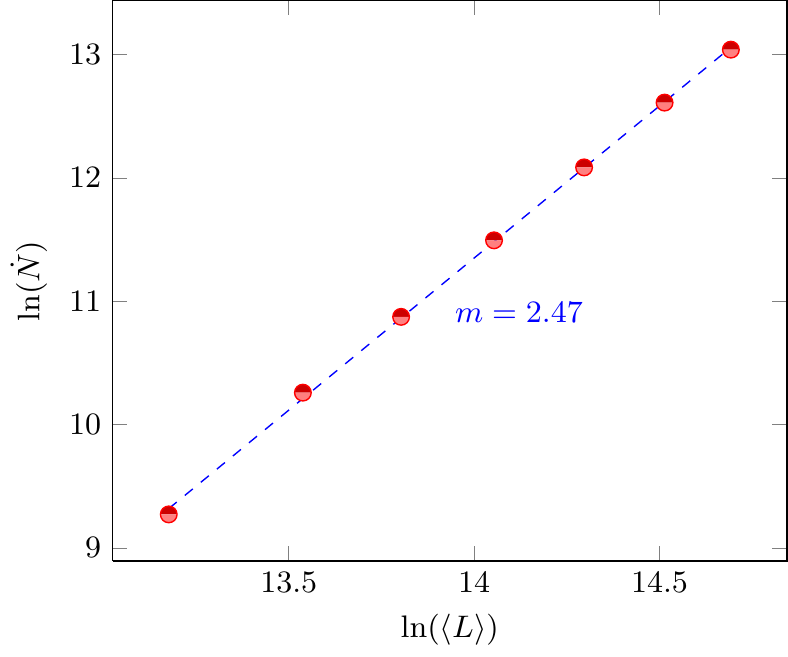}
	\caption{A plot of $\ln(\dot{N})$ versus $\ln(\langle L\rangle)$, finding
    the exponent of $\dot{N}=\mathrm{d}N/\mathrm{d}t\propto\langle L\rangle^m$,
    $m=2.47$.
	}
	\label{fig:reconnections}
\end{figure}

A final test is the rate of vortex reconnections, an important process
in carrying energy down to smaller length scales.  Previous simulations
and analysis show that the rate of vortex reconnections is related to
the line length density by $\dot{n}=C\kappa\langle L\rangle^{5/2}$,
where $\dot{n}$ is the reconnection rate per unit volume and $C$
is a dimensionless constant with a value of approximately 0.1-0.5
~\citep{tsubota00,barenghi04,nemirovskii06,kondaurova14}.  Once again the
earlier Kondaurova {\em et al.}\cite{kondaurova08} results are outliers,
with an overlarge prefactor of $C=2.47$.  Figure~\ref{fig:reconnections}
gives our data on the vortex reconnection rate for a series of trials. We
find an exponent of 2.47 and $C=0.42$, within the normal range.
Our full Biot-Savart calculation matches previous calculations of homogeneous 
superfluid turbulence through multiple comparisons. 

\section{\label{sec:limbiottorus}Nonlocal Distance, $d_{NL}$}

In this section, we investigate further the finding by Adachi {\em et
al.}\cite{adachi10} that the LIA approximation is inadequate for producing
homogeneous turbulence.  We saw in Figure~\ref{fig:ldensfullbiotvsliatime}
that despite the fact that $L$ retains an approximate $\langle
L\rangle\propto v^2$ relationship \citep{schwarz88,aartsdiss},
and even in a time domain before the open-orbit vortex state
dominates, $L$ still deviates significantly between the LIA and
full Biot-Savart calculations. The two methods do produce equally
isotropic systems, however.  For certain limited objectives, such
as Schwarz's\citep{schwarz82,schwarz88} efforts to identify the
attributes necessary for a given behavior, the LIA may be an acceptable
approximation.

\begin{figure}
	\centering
	\includegraphics{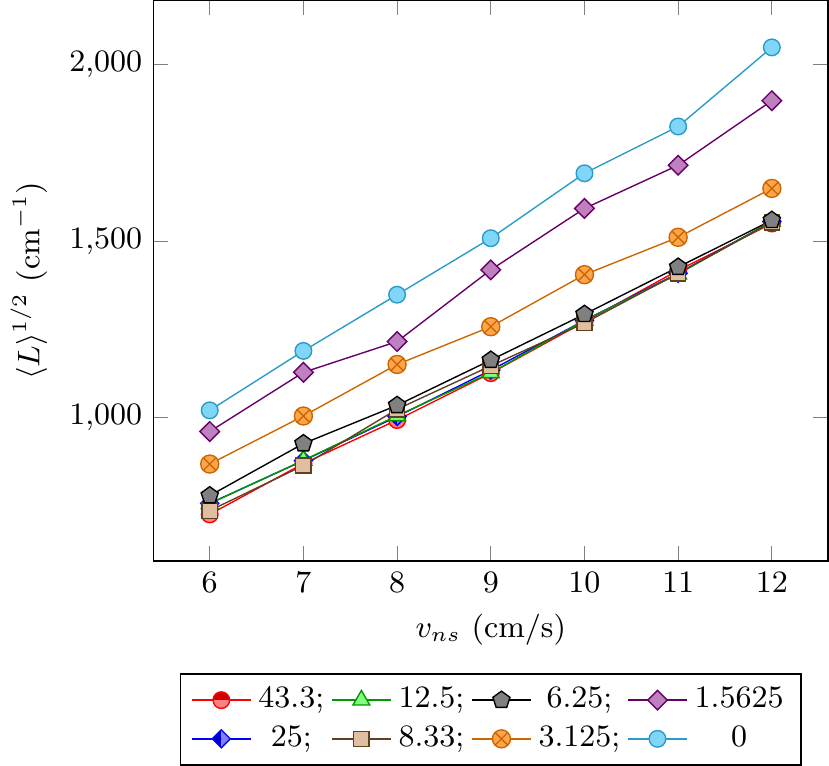}
	 \caption{Effects of nonlocal distance, $d_{NL}$, on 
	the equilibrated line length density, $\langle L\rangle$. Each curve
	is a series of trials at the $d_{NL}$, listed in the legend. The side
	length for the sample region is $D=50 \mu$m, and 
	the largest $d_{NL}$ allows interactions throughout the volume. 
	Averages at each $v_{ns}$ used the initial equilibrated 
	range, similar to Aarts\cite{aartsdiss}.
	}
	\label{fig:ldensvarylimbiot}
\end{figure}

\begin{figure}
	\centering
	\includegraphics{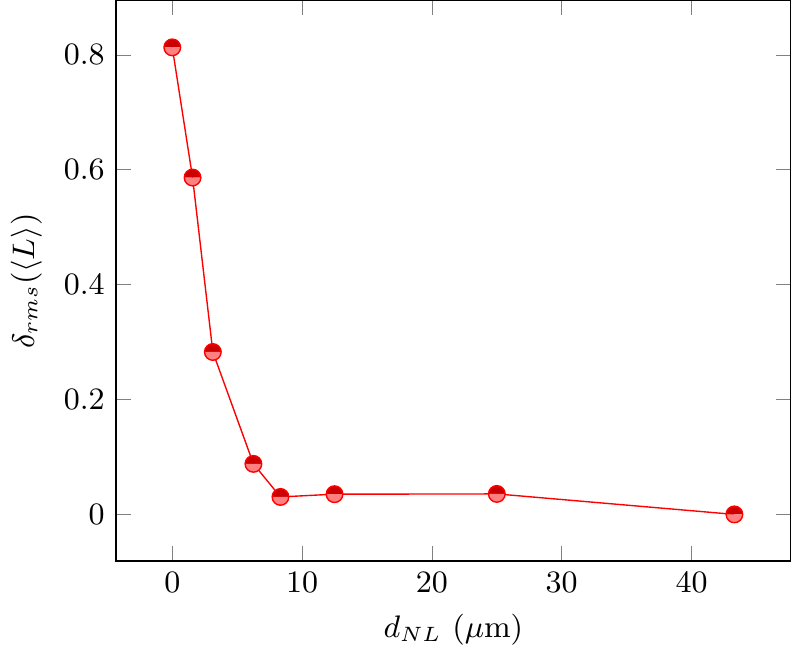}
	 \caption{RMS fractional deviation of 
	 $\langle L \rangle$ from the full Biot-Savart calculation 
	 for several nonlocal distances. Box side is $50 \mu$m.
	 The trials contributing to each point are from
	 Figure~\ref{fig:ldensvarylimbiot}, and use the initial 
	 equilibrated time domain for calculating the average of $L(t)$. 
	 }
	\label{fig:rmsdevldens}
\end{figure}

Here we investigate whether some degree of the computation-saving
power found in the LIA could be retained by truncating the nonlocal
interaction at a distance which we denote by $d_{NL}$.  For each point on
the vortex, we only include contributions to the Biot-Savart integral,
Equation~\ref{eq:toruseqofmotion}, from vortex segments that are within
a distance $d_{NL}$ of the evaluation point.  Our results for $\langle
L\rangle^{1/2}(v_{ns})$ are displayed in
Figure~\ref{fig:ldensvarylimbiot}, for a range of $d_{NL}$ values.
For a sample volume with side $D=50 \mu$m, Biot-Savart interactions among
all pairs of vortices occur at $d_{NL}=\sqrt{3}D/2 = 43.3 \mu$m. As
$d_{NL}$ decreases from this maximum value, the data at first remain
remarkably similar to those using the maximum $d_{NL}$. Eventually the
data begin to deviate, with the line-length density increasing above its
value from the full calculation. The onset of this divergence is even more
apparent in Figure~\ref{fig:rmsdevldens}, which plots the root-mean-square
(RMS) deviation of $\langle L\rangle$ from the full Biot-Savart value:
\begin{equation}
	\delta_{rms}(\langle L\rangle)
	=\left\langle \left( \frac{\langle L\rangle-\langle L_{BS}\rangle}
	{\langle L_{BS}\rangle}\right)^2 \right\rangle_{v_{ns}}^{1/2}.
	\label{eq:rmsdev}
\end{equation}
Here $\langle\ldots\rangle_{v_{ns}}$ represents the average over trials
with different $v_{ns}$ values but the same $d_{NL}$.  The quantity in
Equation~\ref{eq:rmsdev} acts as a measure of the error in our calculation
due to the truncated nonlocal interaction.  Figure~\ref{fig:rmsdevldens}
shows how this quantity varies with $d_{NL}$.
Truncated nonlocal distances of at least $8.33 \mu$m show good agreement
($\lesssim5\%$) with the full Biot-Savart calculation, but a clear
departure arises by $6.25 \mu$m.

\begin{figure}
	\centering
	\includegraphics{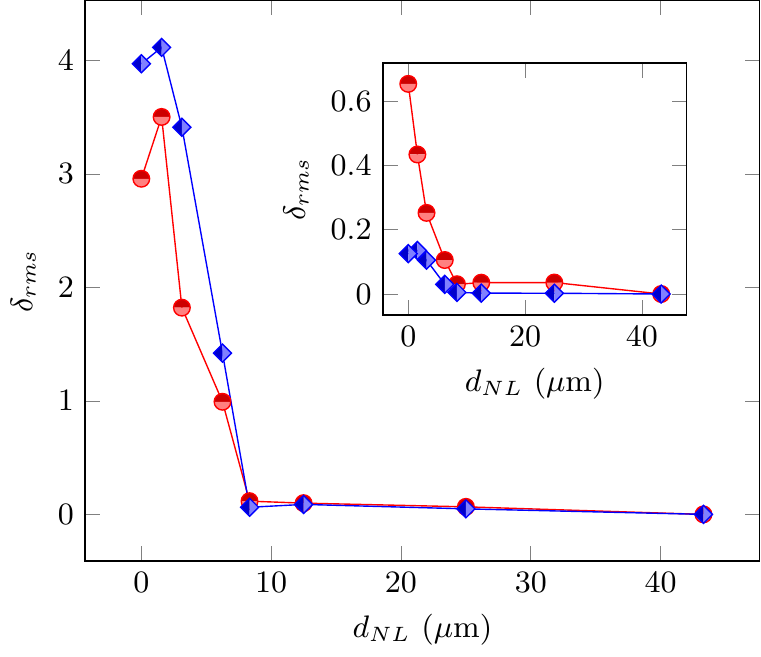}
	 \caption{RMS fractional deviation of 
	 the standard deviations $\sigma(L(t))$ (red circles) 
	 and $\sigma(I_\parallel(t))$ (blue diamonds)
	from the full Biot-Savart calculation plotted versus 
	nonlocal distance. Inset: RMS fractional deviation of the 
	averages $\langle L\rangle$ (red circles) 
	and $\langle I_\parallel\rangle$ (blue diamonds) from 
	the full Biot-Savart calculation plotted versus nonlocal distance.
	Averages use the entire equilibrated time domain. 
	Comparing $\langle L\rangle$ in the inset here
	to the data in Figure~\ref{fig:rmsdevldens} shows the effect of
	changing the time domain used for the averaging.
	}
	\label{fig:rmsdevallparams}
\end{figure}

These time averages all use the initial equilibrated time domain before the
characteristic drop in $L(t)$, seen in 
Figure~\ref{fig:ldensfullbiotvsliatime}, i.e. before the degeneration into the
open-orbit state occurs at the lower $d_{NL}$ values. However, the subjectivity
in selecting the time domain over which to average $L(t)$ could potentially
affect the $d_{NL}$ needed for agreement with the full Biot-Savart calculation.
An alternative approach, for conditions that result in the open-orbit state, is
to calculate the standard deviation of $L(t)$ over the entire time domain after
the initial equilibration. The drop in $L(t)$ upon entering the open-orbit
state sharply increases the standard deviation. Similarly, since from 
Figure~\ref{fig:ldensfullbiotvsliatime}, $I_\parallel$ only deviates from the
full Biot-Savart value outside of the initial equilibrated range, we can also
use this quantity and its standard deviation as indicators of an adequate
$d_{NL}$. There is still a limited sample of data, since we cannot run trials
forever, and we still have to pick out a time when the data start to
equilibrate. Figure~\ref{fig:rmsdevallparams} shows the RMS fractional
deviation of $\sigma(L(t))$, $\sigma(I_\parallel(t))$, $\langle L\rangle$, and
$\langle I_\parallel\rangle$ (making the appropriate substitutions into
Equation~\ref{eq:rmsdev}) for the trials from Figure~\ref{fig:rmsdevldens}
using the entire equilibrated time domain.

From the inset of Figure~\ref{fig:rmsdevallparams} we see
that the critical $d_{NL}$ value is $\lesssim 8.33 \mu$m, the
same whether we use the entire equilibrated time domain in the
averages of $L(t)$ and $I_\parallel(t)$ or simply the initial
time domain, like in Figure~\ref{fig:rmsdevldens}. The main part of
Figure~\ref{fig:rmsdevallparams} shows that the standard deviations of
$L(t)$ and $I_\parallel(t)$ are very good parameters for measuring this
critical $d_{NL}$, showing a pronounced increase in error for nonlocal
distances below this same critical value. We can use a small fraction of
the nonlocal interaction volume and still get a very good approximation
of the Biot-Savart integral for homogeneous superfluid turbulence.

\begin{figure}
	\includegraphics{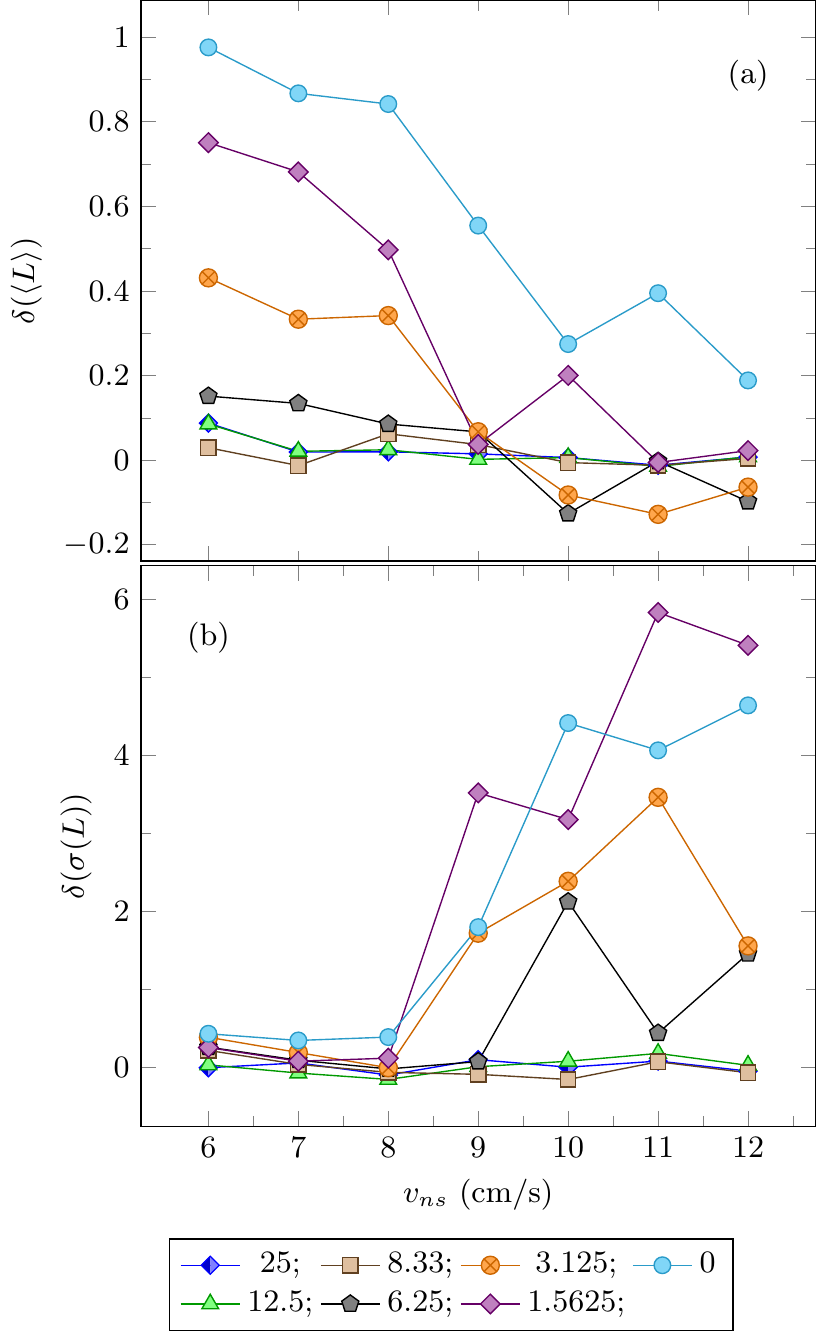}
	   \caption{Fractional deviation of
	   (a) $\langle L\rangle$, and (b) $\sigma(L(t))$, 
	   from the full Biot-Savart calculation plotted versus 
	 driving velocity. Each curve is a different 
	nonlocal distance, $d_{NL}$, indicated by the legend. 
	Averages use the entire equilibrated time 
	domain. 
	}
	\label{fig:devvsvelocity}
\end{figure}

While Figures \ref{fig:rmsdevldens} and \ref{fig:rmsdevallparams} show
averaged results over trials at several values of driving velocity,
Figure~\ref{fig:devvsvelocity} breaks out the behavior of line length
as a function of driving velocity for several nonlocal distances $d_{NL}$.
For a fixed $d_{NL}$, deviations of $\langle L\rangle$ from the full nonlocal
calculation are more pronounced at the lowest velocities. At high velocity
$\langle L\rangle$ is often an average of an initial equilibrated level above
that of the full Biot-Savart calculation, and a later level below the
Biot-Savart value. The overall $\langle L\rangle$ depends on the amount of time
spent in each regime and has little real significance; it may even be
coincidentally close to the correct value. For $\sigma(L(t))$ the reverse is
true, with deviations larger at higher velocities. The slight enhancement of
$\langle L\rangle$ in the initially equilibrated state has little effect on
$\sigma(L(t))$, but the collapse to the open-orbit state drastically increases
its value.  $\langle I_\parallel\rangle$ and $\sigma(I_\parallel(t))$, not shown
in the figure, also deviate more at higher velocities.  As shown in
Figure~\ref{fig:ldensfullbiotvsliatime}~(b), $\langle I_\parallel\rangle$
matches that of the full Biot-Savart calculations until onset of the open-orbit
state, which occurs only for our higher velocities. The abrupt shift in $\langle
I_\parallel\rangle$ also increases $\sigma(I_\parallel(t)$. Because of the
different aspects measured by these quantities, calculating several of them is
useful for finding any discrepancies from the full Biot-Savart calculations.
From a practical perspective, regardless of the underlying cause of any
difference, we want to choose an interaction distance such that every driving
velocity we use produces turbulent behavior matching the full Biot-Savart law.
Our previous averages over all velocities at one $d_{NL}$ pick out deviations at
any $v_{ns}$, and larger number of trials contributiong to each curve reduces
the scatter so that the critical interaction distance becomes clear. 

\begin{figure}
	\includegraphics{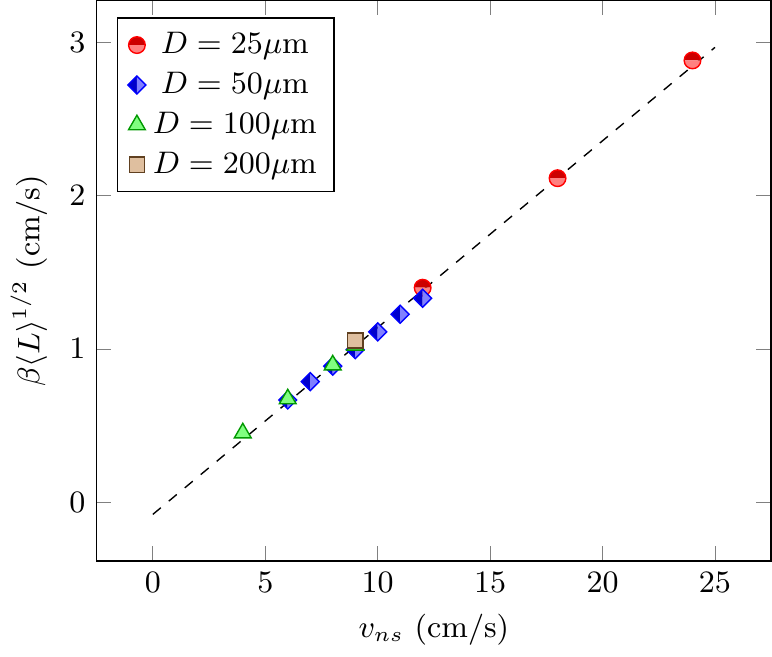}
         \caption{$\beta\langle L\rangle^{1/2}$ data 
	for full Biot-Savart calculations at several system sizes $D$.
	Averages use the entire equilibrated time domain.
	}
	\label{fig:slopeldensallDvals}
\end{figure}

\begin{figure}
	\includegraphics{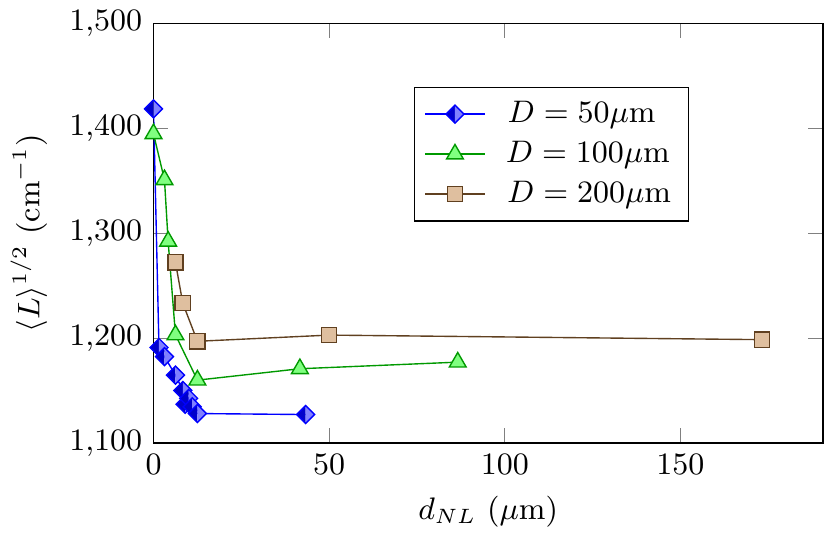}
         \caption{Square root of line length density as function of
	 interaction distance $d_{NL}$, for three box sizes $D$. In each
	 case $\alpha=0.1$ and $v_{ns}=9$ cm/s.
	}
	\label{f:boxsize}
\end{figure}

We next examine how this critical interaction distance varies with system size.
We simulated turbulence in three other system sizes, with other parameters held
constant. $\langle L\rangle$ data for all three 
system sizes are shown in Figure~\ref{fig:slopeldensallDvals}.
The good agreement in $\langle L\rangle$ for 
different system sizes is another mark of homogeneity. Since 
$\langle L\rangle$ is an intensive quantity, with a homogeneous system  it must
be independent of system size. Figure \ref{f:boxsize} illustrates the line
length in the initial equilibrated region, for the three sizes tested at applied
velocity of 9 cm/s. The line length density changes slightly with cell size, but
for a given cell size it remains constant for interaction distances at least
$12.5 \mu$m. For a fixed velocity, the necessary interaction distance does not
depend on system size.

\section{\label{sec:discussion}Discussion}

Similar calculations with other friction coefficients $\alpha$ and applied
velocities $v_{ns}$ shed light on why the non-local interaction need only be
retained for a portion of the cell volume. Our trials with smaller $\alpha$ are
related to lower temperatures, although an attempt to model real behavior of
superfluid helium for these values of $\alpha$ would have to include the
additional $\alpha^\prime$ term that we neglect. The critical interaction
distance is as low as $4 \mu$m, for the lowest friction and highest velocity. We
unfortunately cannot keep $v_{ns}$ entirely fixed while varying $\alpha$. At the
velocities of 9-11 cm/s used in much of this work, turbulence is unsustainable
for $\alpha=0.01$, when less energy is drawn from the driving velocity field. On
the other hand, velocities of 20 cm/s or 50 cm/s, which work for $\alpha=0.01$,
would take impractically long with $\alpha=0.1$.

\begin{figure}
	\includegraphics{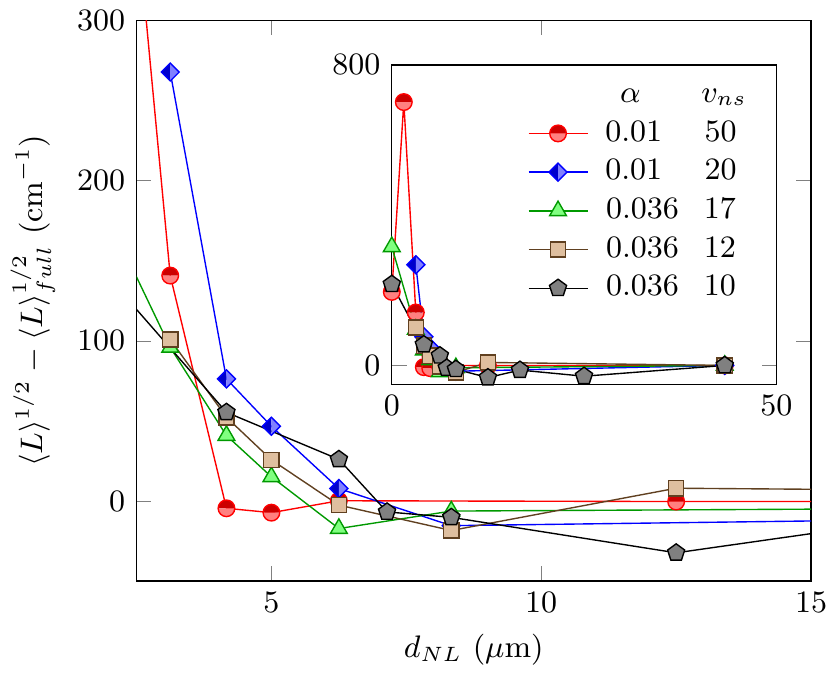}
         \caption{Excess line length as a function of interaction distance, for
	 several combinations of friction parameter $\alpha$ and applied
	 velocity $v_{ns}$. For each curve $\langle L\rangle^{1/2}$ for the
	 largest $d_{NL}$ is subtracted. Main graph: expanded view near critical
	 interaction distance. Inset: entire range. The legend identifies each
	 curve by $\alpha$ and $v_{ns}$, the latter measured in cm/s.
	}
	\label{f:lowTall}
\end{figure}

Figure \ref{f:lowTall} shows results from several trials with lower
friction coefficient. The critical interaction distance changes noticeably
among the trials, generally decreasing as $v_{ns}$ increases. We
suggest that a key issue is the relationship of $d_{NL}$ to the typical
intervortex spacing. Clearly the non-local term has minimal effect
if the calculation volume includes only a single, relatively straight
vortex segment along which the velocity field is being calculated. To
significantly alter the LIA simulation, $d_{NL}$ must be larger than
either the typical vortex separation or the typical vortex radius of
curvature. In the former case, the Biot-Savart calculation would add an
interaction between neighboring vortices too distance for an immediate
reconnection. In the latter case, the time development of a single
contorted filament would change. This is probably most relevant for
highly curved vortices, where self-reconnections are especially likely.

The typical vortex separation $\ell$ can be estimated from the line length
through $\ell=\langle L\rangle^{-1/2}$. For straight, parallel vortices
in a square lattice, this formula gives exactly the nearest-neighbor
separation. The calculation is not exact for other lattices, let alone
for a tangle of curved vortices, but $\ell$ does give a sense of the
intervortex distance.

Table \ref{t:critdist} shows the critical interaction distance along with
the typical vortex separation and the average radius of curvature. The
critical $d_{NL}$ does seem related to $\langle R\rangle$. In fact, the
turn-ups for the different curves of Figure \ref{f:lowTall} are ordered
exactly by the $\langle R\rangle$ values. The relationship to $\ell$ is
less clear. This is hardly surprising, given that the relation of $\ell$
to the nearest-neighbor vortex spacing is inexact and depends on features
like the typical radius of curvature in the tangle. In any case, the
necessary $d_{NL}$ apparently has the same order of magnitude as $\ell$.

\begin{table}
\label{t:critdist}
\begin{tabular}{|lcccc|}
\hline
$\alpha$\hspace*{.45in} & $v_{ns}$ (cm/s)\hspace*{.15in} & critical $d_{NL}$ &
$\hspace*{.2in}\ell$\hspace*{.2in} & \hspace*{.1in}$\langle R\rangle$\hspace*{.1in}\\
\hline
0.01 & 50 & 4 & 5.8 & 3.7\\
0.01 & 20 & 6 & 13.1 & 6.6\\
0.036 & 17 & 6 & 7.5 & 4.8\\
0.036 & 12 & 6 & 10.6 & 6.3\\
0.036 & 10 & 7 & 12.5 & 7.5\\
0.1 & 9 & 9 & 8.9 & 6.3\\
\hline
\end{tabular}
\caption{Onset of deviation from non-local calculation, compared to the
typical intervortex distance and the average radius of curvature. All
lengths are given in $\mu$m. The critical $d_{NL}$ is identified as the
smallest $d_{NL}$ with results indistinguishable by our measures from
the full Biot-Savart calculation. In each case we used 1 $\mu$m steps
in $d_{NL}$ near the critical value.}
\end{table}

Finding the critical $d_{NL}$ comparable to the typical vortex separation
suggests that the key effect of the nonlocal interaction is to draw
together nearby vortex segments until they are within the reconnection
distance. In the LIA, neighboring vortices are invisible to each
other unless they happen to drift close enough to reconnect. If an
extra, small attraction from their Biot-Savart interaction results
in a reconnection that would not otherwise have occured, it has a
significant impact on the subsequent motion.  This finding makes sense
in light of the open-orbit problem. An occasional open-orbit vortex
need not doom a simulation if further reconnections free the vortex
from its open-orbit state.  Only a collective alignment of open-orbit
vortices persists indefinitely.  The occasional rotation step that
Schwarz\cite{schwarz88} used to sustain a vortex tangle operates exactly
by ensuring reconnections. Our findings explain why the non-local
calculation prevents the simultaneous degeneration of vortices into the
open-orbit state. As Nemirovskii notes\cite[p. 152]{nemirovskiireview},
the addition of non-local contributions cannot prevent the reconnections
that lead to open-orbit vortices. Rather, the non-local term enables
{\em additional} reconnections that disrupt the system from settling
into a collective open-orbit configuration.

Beyond the intervortex spacing, the nonlocal contribution is apparently
negligible. This is consistent with the rapid fall-off of the Biot-Savart
law. In addition, for a random tangle the number of vortex segments a
given distance from a calculation point increases as the square of the
distance, and the multiple interactions partially cancel, further reducing
the total. The unimportance of distant non-local terms has been remarked
on previously. A complete Biot-Savart calculation with periodic boundary
conditions would require tiling space with the computational volume
and adding contributions from each copy of the tangle.  Although Ewald
summation provides a rapidly-converging technique \cite{toukmaji}, it
has not been applied to vortex-filament simulations. However, Kondaurova
{\em et al.}\cite{kondaurova14} find that surrounding the main region
with 26 replicas has negligible effect on the results. Our own ``full"
Biot-Savart calculation uses the main volume, with just a few of the
closest image vortices included as well. While it has been standard
practice to exclude very distant contributions, we now argue that
much of the contribution from within the main volume can also be safely
omitted. Yet another effort, in a similar direction, is Baggaley's recent
tree method\cite{baggaley12tree}. Baggaley essentially averages groups
of more distance vortex segments before evaluating the Biot-Savart law,
and finds good agreement with the full Biot-Savart calculations of Adachi
{\em et al.}\cite{adachi10}.  This is to be expected since Baggaley's
method has minimal averaging for very close vortex segments. Those are
often treated the same as in the Adachi calculation. Averaging over more
distant segments has a negligible effect on the statistical properties
of the tangle, as does our truncating the Biot-Savart law to omit these
segments entirely.

Reducing the interaction distance to such a degree can provide
considerable savings in computation time. With interactions omitted
beyond $8.33 \mu$m in a box of side $50 \mu$m, the interaction volume
for each vortex segment is about $2\%$ of the entire box: when the
system is homogeneous, each segment will have only $1/50$ of the other
vortex segments contributing to its nonlocal velocity.  Since other
calculations must be performed for each point in the tangle, the end
result is that simulations with the reduced interaction distance take
$1/3$ to $1/10$ of the computation time. The savings could be greater for
larger computational volumes or higher driving velocities. A question of
practical interest is how small an interaction distance can safely be
used. Our results suggest that ensuring interactions from neighboring
vortices should be adequate. Furthermore, a possible $d_{NL}$ need not
be tested against the full Biot-Savart calculation. It is enough to
obtain equivalent results from a few trials with different $d_{NL}$,
all of which can be small compared to the entire cell size.

\begin{figure}[b]
	\includegraphics{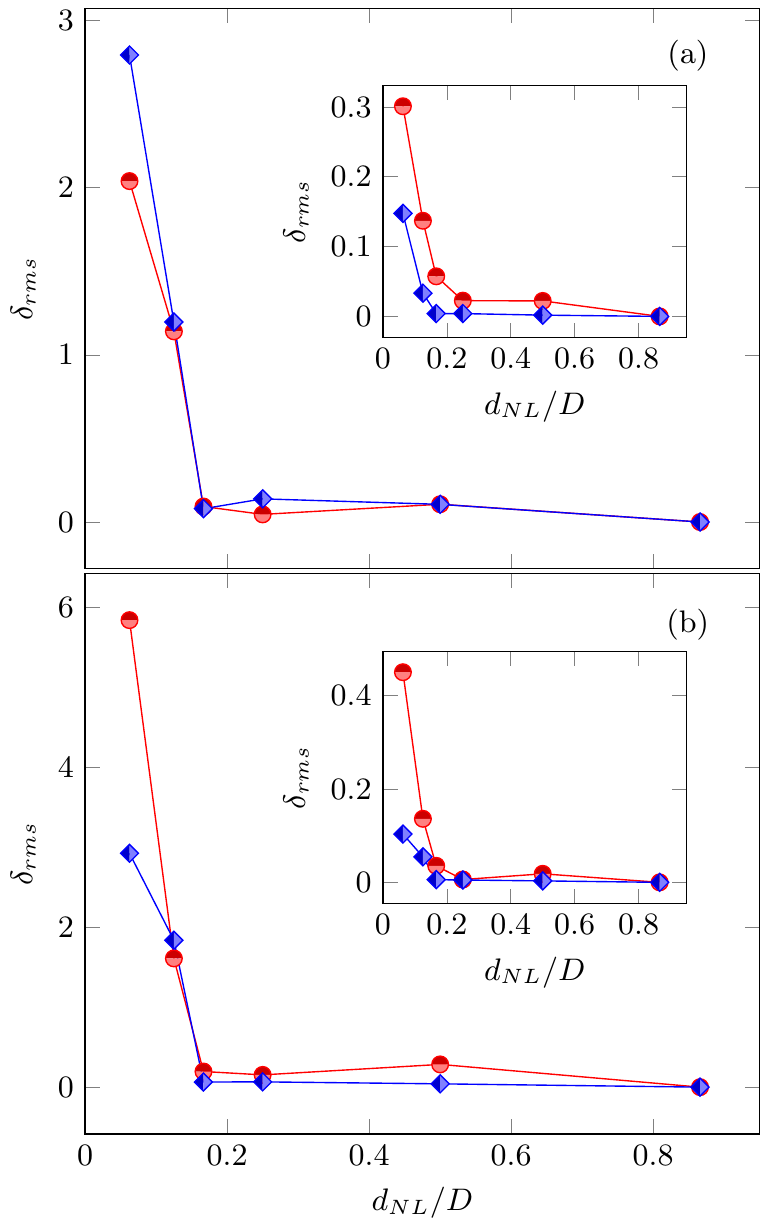}
	\caption{For comparison with Figure~\ref{fig:rmsdevallparams}:
	Figure (a) has system size $D=25 \mu$m, Figure (b) 
	has $D=100 \mu$m. RMS fractional deviation 
	from the full Biot-Savart calculation of the standard deviations
	$\sigma(L(t))$ (red circles)
	and $\sigma(I_\parallel(t))$ (blue diamonds), plotted versus 
	nonlocal distance. Insets: RMS fractional deviation from 
	the full Biot-Savart calculation of $\langle L \rangle$ (red circles)
	and $\langle I_\parallel\rangle$ (blue diamonds), plotted versus
	nonlocal distance.
	Time averages use the entire time domain after initial equilibration.
	}
	\label{fig:rmsdevdiffr0}
\end{figure}

As noted previously, in many cases a single $d_{NL}$ may be desired
across a range of velocities. Usable velocities vary with cell size,
as shown in Figure \ref{fig:slopeldensallDvals}. When the velocity is
too low, tangles are not self-sustaining. When it is too high, the total
vortex length and hence the computational time become prohibitive. As
a practical matter, the necessary interaction distance then becomes
dependent on cell size as well. For the present calculations with
$\alpha=0.1$, we find that $d_{NL}=D/6$ is appropriate for several cell
sizes. Figure \ref{fig:rmsdevdiffr0} illustrates how deviations from the
full non-local calculation begin for two cell sizes, with the velocity
ranges of Figure \ref{fig:slopeldensallDvals}.

\section{\label{sec:conclusion}Conclusion}

Adachi {\em et al.}\cite{adachi10} show the importance of the nonlocal
interaction for homogeneous superfluid turbulence behavior, especially
when simulating this system using periodic boundaries. By using a
truncated Biot-Savart integral, we have found it possible to regain much
of the time saved with the localized induction approximation while still
accurately modeling the statistical behavior of the system. We find that
the nonlocal interaction is only important up to distances of about the
intervortex spacing.

Finally, we address the longstanding question of whether the trouble
with obtaining reproducible results on vortex tangles lies with the LIA,
the reconnection procedure, or the periodic boundary conditions. Our
answer is that the LIA is the main culprit. Its sicknesses cause
unusual sensitivity to other details of the simulation, as shown
previously\cite{baggaley12recon} in the case of reconnection requirements.
Worse, its results can deviate quantitatively from those of other
techniques, even when no qualitatively obvious problem arises. However,
the reason behind the problems with the LIA is primarily its inaccurate
treatment of neighboring vortices, rather than the lack of stretching
that causes difficulty for classical fluid simulations. An appropriate
admixture of a limited non-local term appears to resolve the trouble
with the pure LIA.

\underline{Acknowledgement:} One of us (O. M. Dix) acknowledges support
from the U.S. Department of Education through a GAANN fellowship.

\bibliography{all}
\end{document}